\documentclass[]{article}
\usepackage[normalem]{ulem}
\usepackage{color,colortbl}

\usepackage[%
    backend=bibtex,%
    style=nature,%
    sorting=none%
]{biblatex}
\usepackage{tabularx}
\usepackage{multirow}
\usepackage[utf8]{inputenc}
\usepackage[english]{babel}
\usepackage[top=2cm, left=2.5cm, right=2.5cm]{geometry}
\usepackage{microtype}
\usepackage{authblk}
\usepackage{amsmath}
\usepackage{amssymb}
\usepackage{graphicx}
\usepackage{color,colortbl}
\usepackage{subfigure}
\usepackage{booktabs} 
\usepackage[dvipsnames]{xcolor}
\usepackage{bm}
\usepackage{bbm}
\usepackage{dsfont}
\usepackage{float}
\usepackage{soul}

\usepackage{hyperref}
\usepackage{url}

\usepackage{csquotes}

\usepackage{nameref}

\usepackage{relsize}

\usepackage{stackengine}
\usepackage{scalerel}
\def\rlwd{.4pt}
\def\rlht{1.1pt}
\def\shatvrule{\rule{\rlwd}{\rlht}}
\def\shat#1{%
 \ThisStyle{%
  \setbox0=\hbox{$\SavedStyle#1$}%
  \stackon[0pt]{\stackon[1pt]{\ensuremath{\SavedStyle#1}}{%
    \shatvrule\kern\wd0\kern-\rlwd\kern-\rlwd\shatvrule}}%
    {\rule{\wd0}{\rlwd}}%
 }%
}

\title{Beyond spiking networks: the computational advantages of dendritic amplification and input segregation}

\author [1, *, +] {Cristiano Capone}
\author [1, *] {Cosimo Lupo}
\author [2] {Paolo Muratore}
\author [1] {Pier Stanislao Paolucci}

\affil [1] {INFN, Sezione di Roma}
\affil [2] {SISSA, Trieste, Italy}
\affil [*] {These authors contributed equally to this work}
\affil [+] {cristiano0capone@gmail.com}

\date{}

\begin{document}

\maketitle

\definecolor{cristianoorange}{rgb}{0.9,0.3,0.}
\definecolor{minoblue}{rgb}{0.,0,0.8}
\definecolor{minogreen}{rgb}{0.565,0.933,0.565}
\definecolor{paoloviolet}{rgb}{0.788, 0.36, 0.92}
\definecolor{pierred}{rgb}{0.9, 0.1, 0.1}

\newcommand{\cristiano}[1]{{\bf\color{cristianoorange}{#1}}}
\newcommand{\mino}[1]{{\bf\color{LimeGreen}{#1}}}
\newcommand{\stmino}[1]{{\color{LimeGreen}{\st{#1}}}}
\newcommand{\paolo}[1]{{\bf\color{paoloviolet}{#1}}}
\newcommand{\pier}[1]{{\bf\color{pierred}{#1}}}

\newcommand{\lock}[0]{
\mbox{\scriptsize$\begin{smallmatrix}\mathsmaller{\bm{\cap}} \\[-1pt] \mathlarger{\blacksquare} \end{smallmatrix}$}
}

\vspace{-10mm}
\begin{abstract}

The brain can efficiently learn a wide range of tasks, motivating the search for biologically inspired learning rules for improving current artificial intelligence technology.
Most biological models are composed of point neurons, and cannot achieve the state-of-the-art performances in machine learning.
Recent works have proposed that segregation of dendritic input (neurons receive sensory information and higher-order feedback in segregated compartments) and generation of high-frequency bursts of spikes would support error backpropagation in biological neurons.
However, these approaches require propagating errors with a fine spatio-temporal structure to the neurons, which is unlikely to be feasible in a biological network.

To relax this assumption, we suggest that bursts and dendritic input segregation provide a natural support for biologically plausible target-based learning, which does not require error propagation.
We propose a pyramidal neuron model composed of three separated compartments, the basal one (receiving the sensory input) and the apical ones (receiving the recurrent and the context/teaching signals).
A coincidence mechanism between the basal and the apical compartments allows for generating high-frequency bursts of spikes.
This architecture allows for a burst-dependent learning rule, based on the comparison between the target bursting activity triggered by the teaching signal and the one caused by the recurrent connections, providing the support for target-based learning. We show that this framework can be used to efficiently solve spatio-temporal tasks, such as the store and recall of 3D trajectories.

We argue that learning networks with this architecture enjoy a number of desirable properties, among which the capability to learn without error propagation and the possibility to robustly select context-dependent responses to given sensory stimuli. 
Finally, we suggest that this neuronal architecture naturally allows for orchestrating ``hierarchical imitation learning'', enabling the decomposition of challenging long-horizon decision-making tasks into simpler subtasks. This can be implemented in a two-level network, where the high-network acts as a ``manager'' and produces the contextual signal for the low-network, the ``worker''.

\end{abstract}

\section{Introduction}

Biological networks of neurons can solve a disparate variety of tasks with high energetic and sample efficiency, motivating the search for biologically inspired learning rules for improving artificial intelligence.

The last decades have seen consistent progresses in the development of efficient neural networks (Fig.~\ref{fig1}A), taking more and more inspiration from biology.
The first generation of neural networks was based on perceptrons (also referred to as McCulloch-Pitts neurons or threshold gates), only capable to provide a digital output (as discussed in \cite{maass1997networks}). 
The second generation was based on computational units that apply an ``activation function'' with a continuous set of possible output values to a weighted sum (or polynomial) of the inputs. Typical examples are feedforward and recurrent sigmoidal neural networks.
In the 90s, experimental results from neurobiology led to a third generation of neural network models, employing spiking (or ``integrate-and-fire'') neurons as computational units \cite{maass1997networks, bellec2020, muratore2021target, capone2021error}. 
Networks of spiking neurons are, with regard to the number of neurons that are needed, computationally more powerful than these other neural network models  \cite{maass1997networks}. Moreover, they allow improved energy efficiency, and the possibility to encode information through spike timing. %
Despite these theoretical and technological advancements, most biologically inspired neural networks are composed, so far, of point neurons \cite{nicola2017supervised, bellec2020}, and cannot achieve the state-of-the-art performances of artificial intelligence (e.g., they struggle to solve the credit-assignment problem \cite{payeur2021burst}).

Recent findings on dendritic computational properties \cite{poirazi2020illuminating} and on the complexity of pyramidal neurons dynamics \cite{larkum2013cellular} motivated the study of multi-compartment neuron models in the development of new biologically plausible learning rules \cite{urbanczik2014learning,guerguiev2017towards,sacramento2018dendritic,payeur2021burst}.
These observations are giving rise to a fourth generation of neural networks, composed of spatially extended,  bursting neurons, making it possible to exploit learning paradigms that were not accessible in the previous generations of neurons, as we will demonstrate in this paper.

It has already been proposed that segregation of dendritic input~\cite{guerguiev2017towards} (i.e., neurons receive sensory information and higher-order feedback in segregated compartments) and generation of high-frequency bursts of spikes \cite{payeur2021burst} would support backpropagation in biological neurons. However, current approaches require propagating errors with a fine spatio-temporal structure to all the neurons, and it is not yet clear whether this is possible in biological networks. For this reason, in the last few years, target-based approaches \cite{lee2015difference,depasquale2018full,manchev2020target,meulemans2020theoretical,muratore2021target} started to gain more and more interest.
In a target-based learning framework, the targets --- rather than the errors --- are propagated through the network \cite{lee2015difference,manchev2020target}. In this way, it is possible to directly suggest to the network the internal solution to a task \cite{depasquale2018full, muratore2021target, capone2021error}.
However, the spontaneous activity and the target activity of the network need to be evaluated at the same time~\cite{depasquale2018full, muratore2021target}. This is usually solved by evaluating the two activities in two different networks, which is not natural in terms of biological plausibility.

In the present work, we show that bursts and dendritic input segregation offer a natural solution to this dilemma.
Our learning rule builds upon an important architectural assumption (see Fig.~\ref{fig1}B): the input arriving to the apical dendritic compartments is further segregated in local predictions (to the proximal apical compartment) and teaching/contextual signals (to the apical distal compartment).
A coincidence mechanism between the basal and the apical distal (or the apical proximal) inputs generates a burst \cite{larkum2013cellular}, eventually defining the target (or predicted) spatio-temporal bursting dynamics of the network.



This segregation, besides being a natural way to compare local predictions and higher-order suggestions, can be justified by geometric considerations: as the mentioned signals come from very different spatial locations, it is reasonable to assume that they would arrive in different regions of the neuron.
Our assumption can also be interpreted as a theoretical prediction to be validated by dedicated experiments.

In our model, we exploit dendritic computation to let arbitrary signals act as teaching signals which drive the learning procedure in a biologically plausible fashion. This allows us to flexibly store and recall arbitrary trajectories, with performances that are competitive with the state-of-the-art error-based approaches.
Finally, we will show that this neuronal architecture naturally allows for orchestrating \textit{hierarchical imitation learning}, enabling the decomposition of challenging long-horizon decision-making tasks into simpler subtasks~\cite{le2018hierarchical, pateria2021hierarchical}, through the implementation of a two-level network, with the high-network acting as a ``manager'' and producing the contextual signal for the low-network, the ``worker''.

\section{Results}

\subsection{Target-based learning with bursts}
\label{subsec:target_based}

\subsubsection{Neuronal architecture}

\begin{figure}[h!]
\centering
\includegraphics[width=130mm]{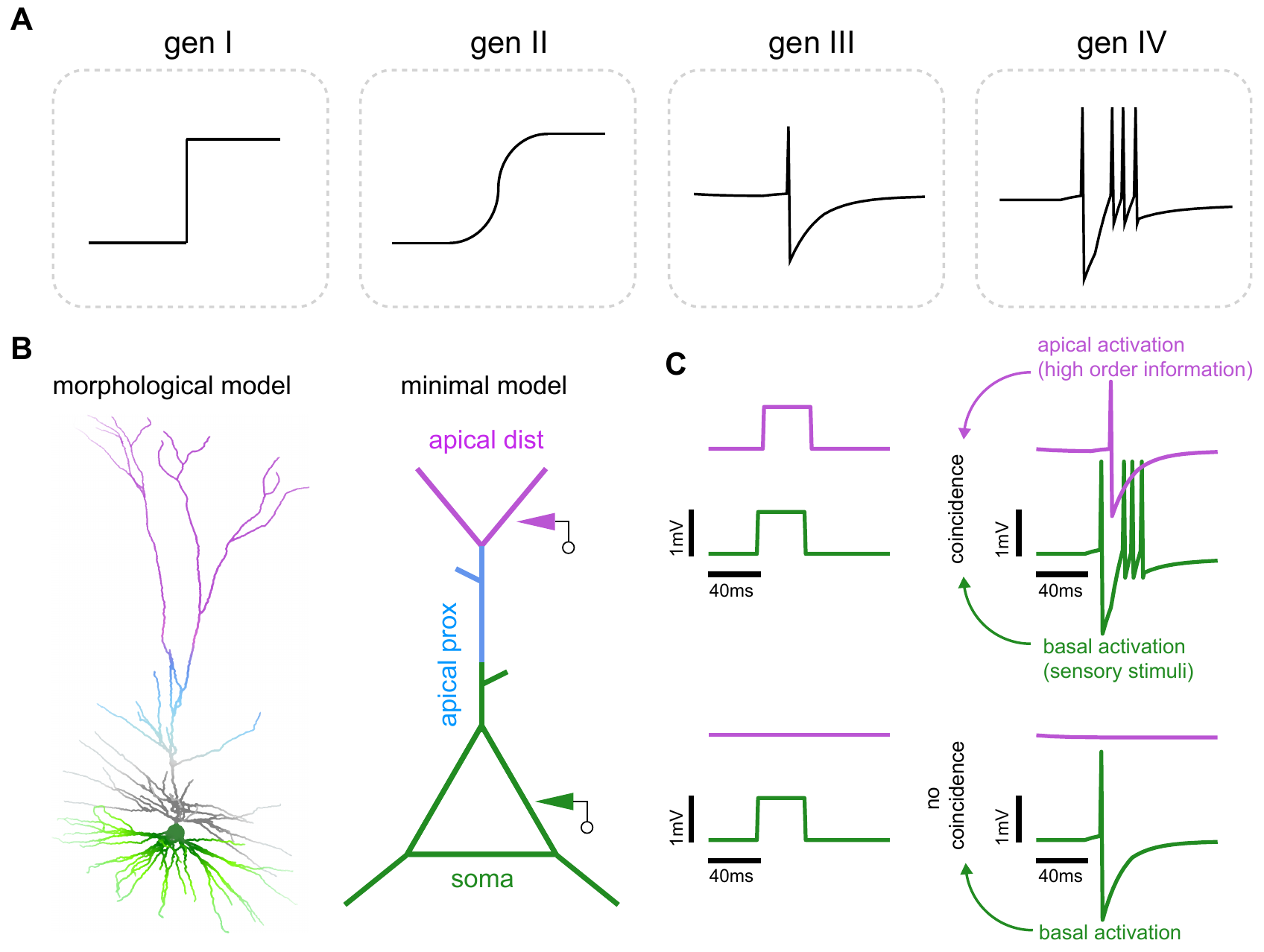}
\caption{
\textbf{The multi-compartment neuron model.} 
\textbf{A}. The four generations of neural networks, from ``threshold gate'' and ``activation function'' models, to spiking neurons and finally to multi-compartment neurons producing high-frequency bursts.
\textbf{B}. (Left) Representation of the morphology of a L5 pyramidal neuron. (Right)  Our three-compartment simplified model of a L5 neuron: the soma (green) receives sensorial inputs; the apical proximal compartment (blue) receives the recurrent connections from the other neurons in the network; the apical distal compartment (purple) receives teaching/contextual signals from other areas of the cortex.
\textbf{C}. The coincidence mechanism implemented thanks to dendritic segregation, taking inspiration from the mechanism shown in \cite{larkum2013cellular}. (Top) When a dendritic spike occurs in coincidence with a somatic spike, a high-frequency burst of somatic spikes is generated. (Bottom) When this coincidence does not occur, only isolated spikes can be generated.
}
\label{fig1}
\end{figure}

Inspired by the morphology of L5 pyramidal neurons (Fig.~\ref{fig1}B, left), we defined a neuron model composed of three separated compartments (Fig.~\ref{fig1}B, right): the basal one (i.e. the soma), receiving the sensorial input; and two apical ones, the proximal apical compartment, receiving recurrent connections from the network, and the distal apical compartment, receiving the context/teaching signal from other areas of the cortex, with a higher level of abstraction.

Each of these compartments is characterized by a membrane potential modeled through a leaky-integrate-and-fire dynamics. The spike emitted by the soma of the $i$-th neuron is described by variable $z_i^t$, which is equal to $1$ when the spike is emitted at time $t$ and $0$ otherwise. The spikes emitted by the proximal and distal apical compartments are then described by variables $a_i^t$ and $a_i^{\star, t}$, respectively. The underlying idea is that the distal compartment provides a target for the proximal one, motivating the use of the superscript symbol $\star$, which indicates the variables concerning the target.

Following \cite{larkum2013cellular}, a coincidence mechanism between the basal and the apical compartments has been implemented, yielding high-frequency bursts of spikes from the soma. In more detail, after a somatic spike, $z_i^t=1$, a coincidence window is opened for a time interval $\Delta T$. This is described by the variable $\shat{z}_i^t$, the indicator function for $t' \in [t, t + \Delta T]$, which is $1$ during this time window and $0$ elsewhere. If a spike is generated by the distal or proximal apical compartment within such time window, $a_i^{t'}=1$ or $a_i^{\star, t'}=1$, with $t'\in[t, t + \Delta T]$, a high-frequency burst of spikes is then produced (Fig.~\ref{fig1}C, top).
Such coincidence mechanism is defined for both the distal and the proximal apical compartments; the functional differentiation between them will be clarified in the following section.
The resulting proximal and distal burst variables can be hence written respectively as:
\begin{align*}
B_{i}^{t+1} &= \shat{z}_i^{t}  \, a_{i}^{t+1}\\
B_{i}^{\star, t+1} &= \shat{z}_i^{t} \, a_{i}^{\star, t+1}
\end{align*}

\subsubsection{Burst-mediated plasticity rule}

\begin{figure}[h!]
\centering
\includegraphics[width=120mm]{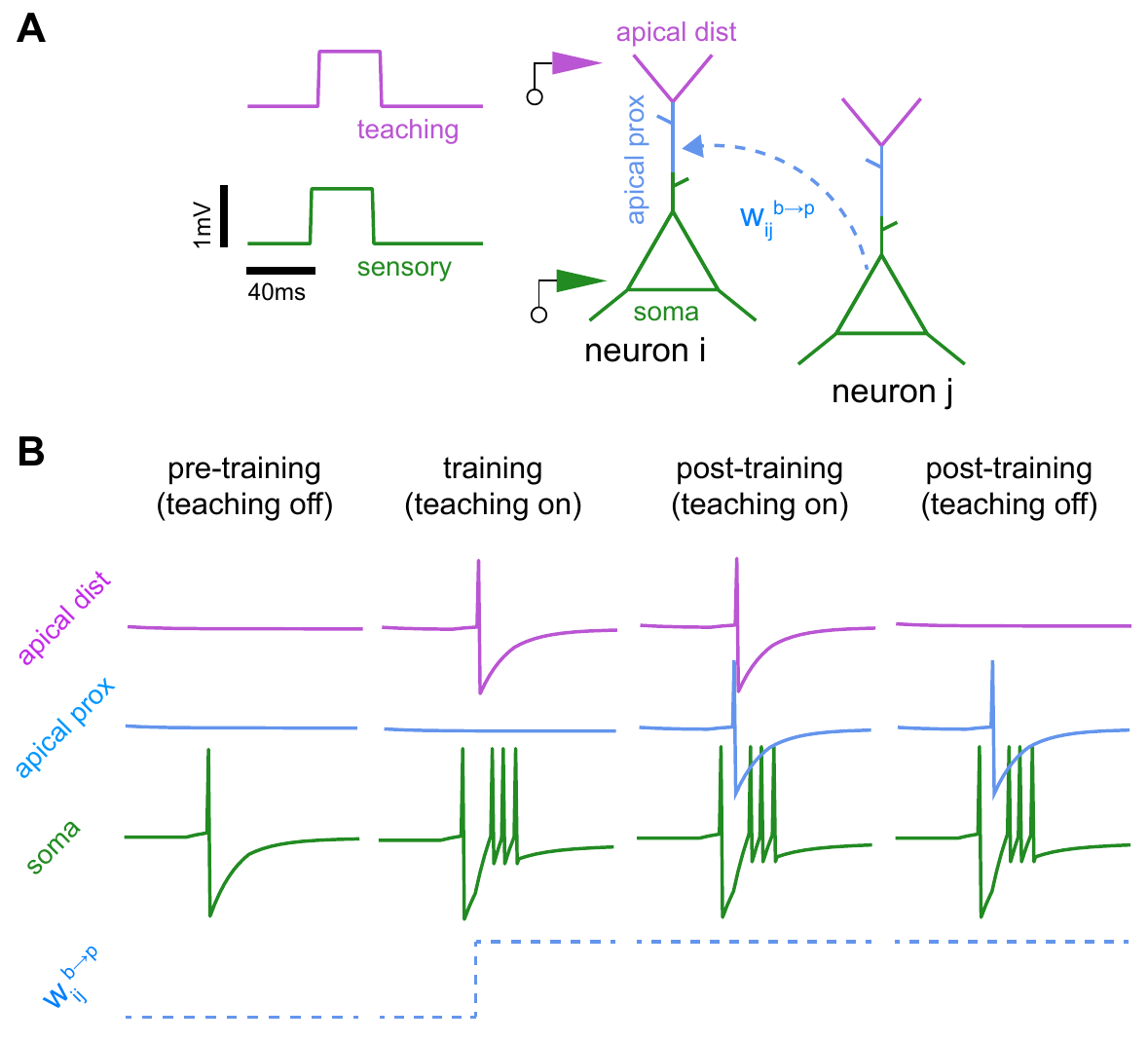}
\caption{
\textbf{Teaching through burst-mediated plasticity rule.}
\textbf{A}. Our three-compartment neuron receive a spatially segregated input:  the somatic compartment (green) receives sensorial inputs; the apical proximal compartment (blue) receives the recurrent connections from the network; the apical distal compartment (purple) receives teaching/contextual signals from other areas of the cortex.
\textbf{B}. Schematics for the teaching process. Time course of the membrane potentials of the three compartments (colors as in panel A) and of a synaptic weight (blue dashed). The plasticity rule we derived, allows for synaptic changes only when a burst occurs. It aims at aligning proximal spikes ($a$) generated by the network, with distal spikes ($a^{\star}$) induced by the teacher/context. 
(First column) Before learning, recurrent connections are not trained and no activity from the apical proximal compartment is detected.
(Second column) The coincidence between a somatic spike (green) and an apical distal spike (purple) triggers a burst (green). However, mismatch between distal (purple) and proximal spikes (blue) when the somatic coincidence window is open, induces the change of pre-synaptic weights during the training phase (dashed blue).
(Third column) After learning, the network is able to self-sustain the apical activity and then correctly reproduce the target determined by the teaching signal.
(Last column) The neuron is capable to produce the learned burst also when the teaching signal is no longer present.
}
\label{fig2}
\end{figure}

The modular architecture of L5 pyramidal neurons, together with dendritic active computation, is thought to have astonishing computational properties \cite{poirazi2020illuminating}. We propose that the multi-modular architecture of our module can be exploited to develop new classes of learning rules.
Starting from general considerations we note how, because of the spatial segregation of the neuron, we can expect that a generic learning rule is a function of the activities of each compartment, and can be described by the expression $f \left(z_i, a_i, a^\star_i \right)$. 
If we request space and time locality, then, we can further refine our description and impose a generic plasticity rule for the recurrent weights $w_{ij}^{b \to p}$ of the form: $\Delta w_{ij}^{b \to p} = f_\mathrm{post} \left(z_i, a_i, a^\star_i \right) g_\mathrm{pre} \left( \bm{\mathrm{s}}_j \right)$, where $f_\mathrm{post}$ and $g_\mathrm{pre}$ are generic functions.

In this work we present a particular instance of this novel class of burst-dependent plasticity rules, which naturally enables target-based learning.
More specifically, we propose that the pattern of bursts defined by the proximal compartment (receiving the recurrent connections $w_{ij}^{b \to p}$ from the network, see Fig.~\ref{fig2}A, blue arrow) should mimic the ones induced by the distal compartment (which receives the teaching signal, see Fig.~\ref{fig2}A, purple triangle), with the sensory input entering the somatic compartment (see Fig.~\ref{fig2}A, green triangle). This is made possible by using the following plasticity rule for recurrent weights $w_{ij}^{b \to p}$ (which can be derived analytically through a likelihood maximization, see Methods for details):

\begin{equation}
    \Delta w_{ij}^{b \to p} = \eta   \left[a_{i}^{\star, t+1} - a_{i}^{t+1} \right] \shat{z}_{i}^t \, e_j^t
    \label{plasticity_rule}
\end{equation}
where $e_j^t = \partial u_i^t/\partial w_{ij}^{b \to p}$ is referred to in the literature as the \textit{spike response function} \cite{urbanczik2014learning}.
In other words, such plasticity rule aims at aligning in time apical proximal spikes with apical distal ones when the somatic window $\shat{z}_i^t$ is open. We remark that such a learning rule can be computed online, and only requires observables which are locally accessible to the synapses in space and time.

These ingredients allow to arbitrarily train a neuron to produce a burst by using the proper teaching signal (see Fig.~\ref{fig2}A). For example, let's consider a neuron that produces a somatic spike in response to a sensory input (see Fig.~\ref{fig2}B, first column). 
To induce this neuron to produce a burst in response to that sensory stimulation, it is sufficient to inject a teaching current  capable to induce a spike in the apical distal compartment and hence a burst in the soma (see Fig.~\ref{fig2}B, second column). The mismatch between the proximal and the distal response (see Fig.~\ref{fig2}B, second column, blue and purple lines) triggers synaptic plasticity, inducing an increase of synaptic weights (see Fig.~\ref{fig2}B, second column, blue dashed line).
After training, the pyramidal neuron is capable to produce a burst (Fig.~\ref{fig2}B, third column) also when the teaching signal is no longer present (see Fig.~\ref{fig2}B, last column), thanks to the proximal apical spike induced by the learned recurrent weights.

An important feature of our model is what we call \textit{teacher neutrality}, i.\,e. the presence of the teaching signal becomes irrelevant after the training (see Fig.~\ref{fig2}B, comparison between third and last column). Indeed, if the apical proximal compartment emits a spike when the somatic window is already open, and a somatic burst is consequently generated, a further apical distal spike does not trigger more bursts. This feature is essential for teacher learning and can be biologically justified thanks to mechanisms of \textit{apical saturation}. 

\subsubsection{Store and recall}

\begin{figure}[H]
\centering
\includegraphics[width=\textwidth]{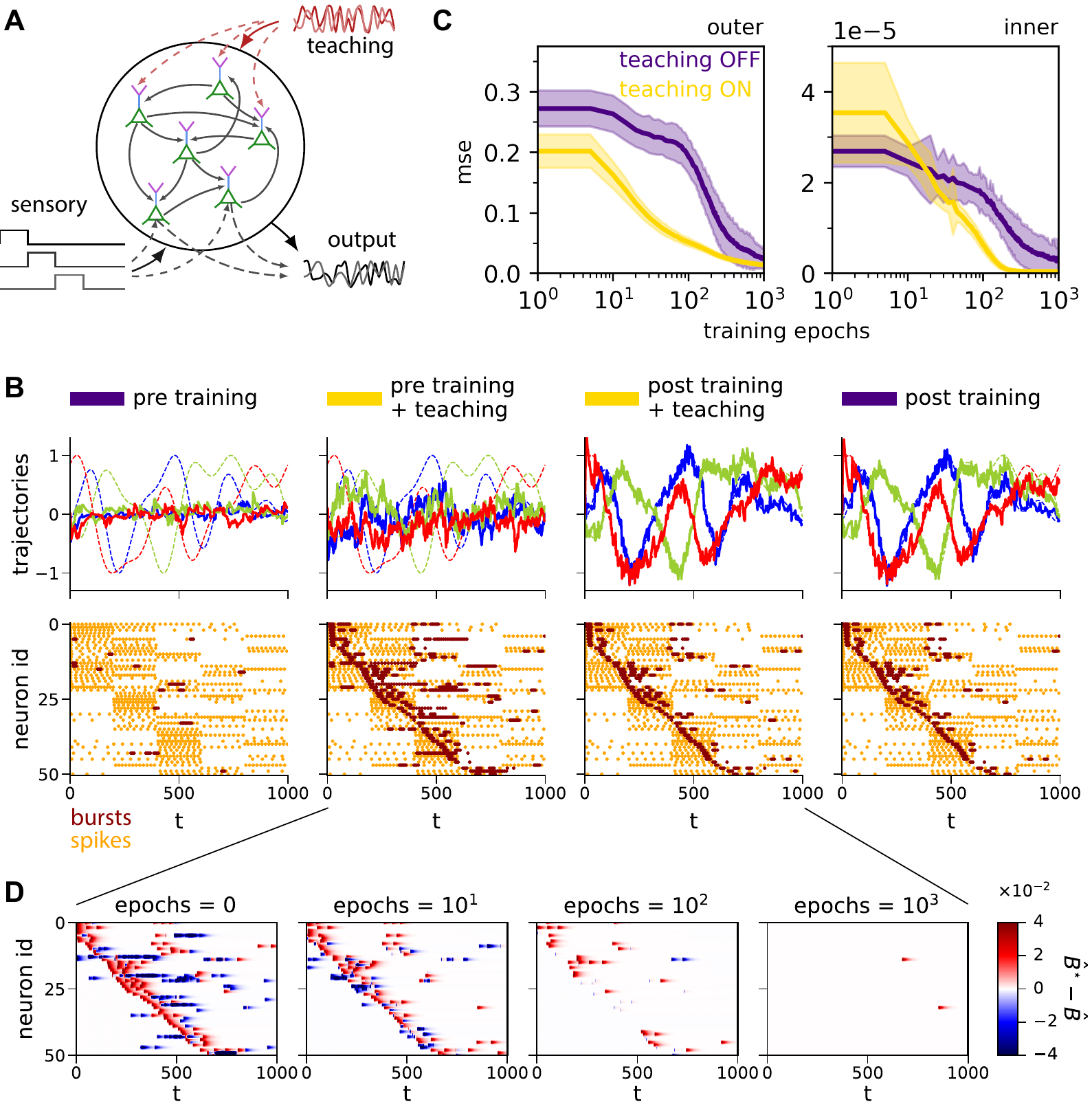}
\caption{
\textbf{Model structure.} 
\textbf{A}. Sketch of the network setting used for the store-and-recall task of a 3D trajectory.
\textbf{B}. For each of the four panels, in the top row we reported the target trajectory (dashed lines) together with the output produced by the network (solid lines); in the bottom row, isolated spikes (yellow) and bursts (brown) from the somatic compartment. Before learning (first column), the network randomly produces patterns of bursts. When the teaching signal is present (second column) the network produces a combination of the initial bursts and the one induced by the teaching signal. After the training procedure (third column), the network learned to produce only the correct bursts. (fourth column) The network is eventually able to autonomously reproduce the proper spatio-temporal pattern of bursts and output, also in the absence of the teaching signal.
\textbf{C}. (Left) mse of the output 3D trajectory against the target one during the training phase with (yellow) and without (violet) the suggesting signal. (Right) mse between the internal pattern of bursts and the one suggested by the teaching signal, with (yellow) and without (violet) the suggesting signal. Averages 50 independent realizations.)
\textbf{D}. Rastergram of the internal local error $\hat{B}_i^{\star, t}  - \hat{B}_i^{t}$ for different training epochs, from the beginning to the end of the teaching procedure.
}
\label{fig3}
\end{figure}

As a first learning instance, we propose the store and recall of a 3D trajectory $y_k^{\star, t}$ ($k = 1,\dots,3$; $t = 1,\dots,T$; $T=1000$) in a network of $N = 500$ neurons (see Fig.~\ref{fig3}A, $400$ bursting neurons with the pyramidal architecture described above, plus $100$ non-bursting point neurons).
We chose $y_k^{\star, t}$ as a temporal pattern composed of $3$ independent continuous signals, each of which specified as the superposition of the four frequencies $f_n \in \left\{1, 2, 3, 5 \right\}$ Hz, with uniformly extracted random phases $\phi \in \left[0, 2 \pi \right]$:
\begin{equation*}
y_k^{\star, t} = \sum_{n=1}^4 A_{k,n}\cos{(2\pi f_n t + \phi_{k,n})} \, , \quad k=1,2,3
\end{equation*}
Amplitudes are randomly extracted as well, $A \in \left[0.5, 2.0 \right]$, eventually normalized in order to have trajectories within the $[-1,1]$ interval.

This target trajectories are randomly projected through a Gaussian matrix with variance $\sigma_{\mathrm{targ}}^2$ to the (distal) apical dendrites of the network as a teaching signal. This input shapes the spatio-temporal pattern of spikes $a_{i}^{\star, t}$ from the distal apical compartment, as well as the related target spatio-temporal pattern of bursts $B_i^{\star, t}$ (Fig.~\ref{fig3}B, bottom, brown points) as described above.

A clock signal serving as a sensory input (see Fig.~\ref{fig3}A)  is randomly projected (through a Gaussian matrix with variance $\sigma_{\mathrm{in}}^2$) to the somatic dendrites. In more detail, the clock is here modeled as a sort of time step function with $I$ steps, such that at each time $t$ only component $i=\lfloor I\cdot t/T\rfloor$ is equal to 1, while others are 0 (see Table~\ref{table1} for model parameters).

Before learning, the network randomly produces a spatio-temporal pattern of bursts, that does not encode any relevant information (see Fig.~\ref{fig3}B, first column, bottom panel). The internal bursting is translated into the output $y$ (see Fig.~\ref{fig3}B, first column, top panel) by means of a read-out matrix $w_{\mathrm{out}}$, randomly initialized and to be trained following the rule derived by minimizing the mean squared error ($\mathrm{mse}$) between the target output and the network output:
\begin{equation}
    \Delta w^{\mathrm{out}}_{ki} = \eta_{\mathrm{out}} \left[y^{\star, t}_k - \sum_{h} w^{\mathrm{out}}_{kh} \hat{B}^t_h\right] \hat{B}^t_i
\end{equation}
where $\hat{B}$ is a time-smoothed version of burst variable $B$ (see Methods for details).

In order to train the network, the teaching signal is given by the target trajectory $y^{\star}$ itself. It influences the network dynamics, producing a pattern of bursts (see Fig.~\ref{fig3}B, second column, bottom) that is a super-position of the ones originally generated by the randomly inizialized network and the ones induced by the teacher.
The resulting mismatch between the network predictions ($B_i^{t}$) and the teacher suggestions ($B_i^{\star, t}$) activates the plasticity of recurrent weights $w_{ij}^{b \to p}$, according to Eq.~\eqref{plasticity_rule}. This mismatch is represented in Fig.~\ref{fig3}D for different training epochs, by means of the rastergram of the internal local error $\hat{B}_i^{\star, t} - \hat{B}_i^{t}$.

At the end of learning, such mismatch is no longer present, and the network only produces the bursts suggested by the teacher (induced by the teaching signal, see Fig.~\ref{fig3}B, third column), encoding the solution of the task. Also, readout weights have been learned, and the output correctly reproduces the target trajectory (see Fig.~\ref{fig3}B, third column, top panel). In other terms, the network is eventually capable to reproduce the same spatio-temporal pattern of bursts and the same output in absence of the teaching signal (see Fig.~\ref{fig3}B, fourth column, top panel). This ability of performing the same dynamics equally in presence or absence of the teaching signal is an outstanding property of our model.






The same task is addressed in \cite{muratore2021target} and \cite{bellec2020}, obtaining $\mathrm{mse}$ values of $0.001$ and $0.01$, respectively. Though the latter result is very similar with the present one (approximately $0.01$, averaged over 50 realizations), a direct comparison is unfair, since the target is here encoded only through the bursts, that are way less than spikes, so providing a much sparser encoding. On the other hand, our model results in a remarkable improvement in terms of biological plausibility.

\subsection{Apical contextual signals to robustly select desired responses}
\label{subsec:context_selection}

\begin{figure}[ht!]
\centering
\includegraphics[width=\textwidth]{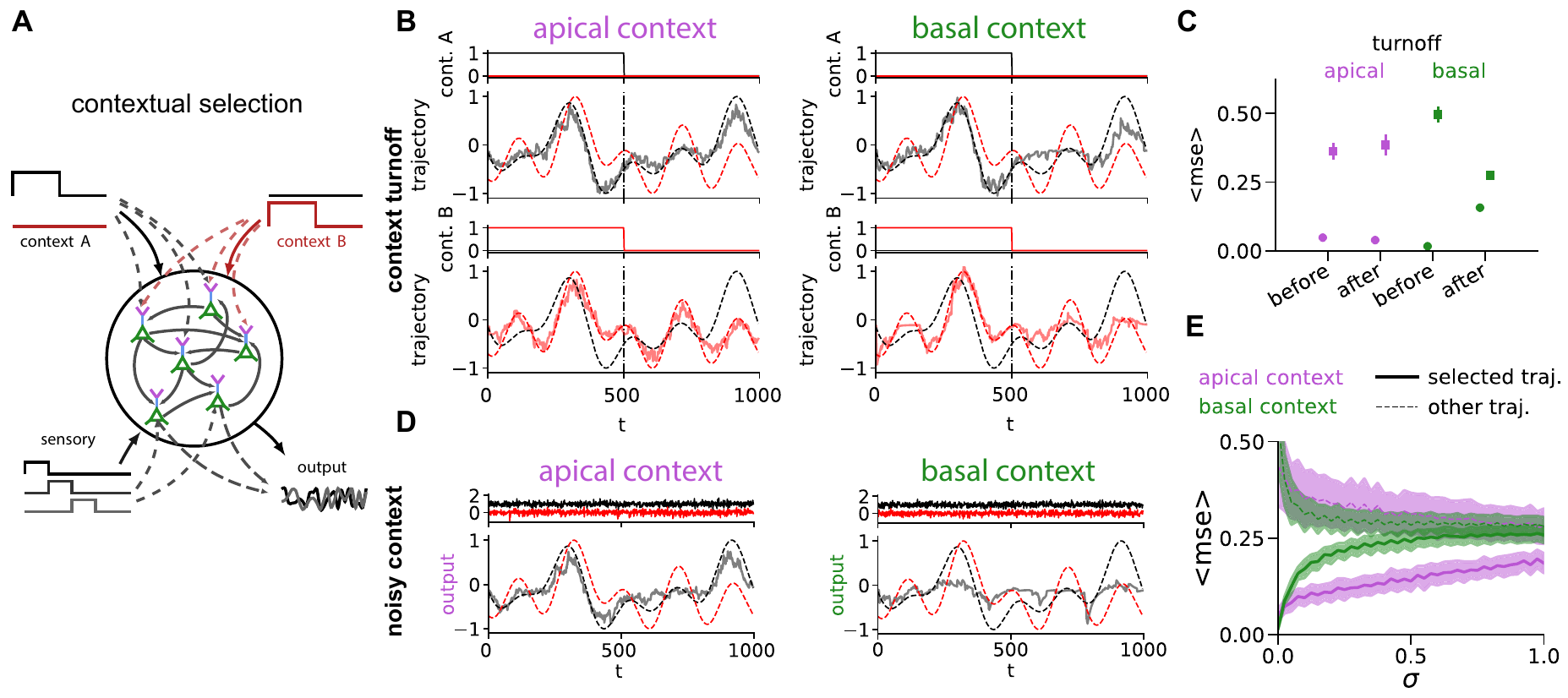}
\caption{
\textbf{Apical signals for contextual selection.}
\textbf{A}. Sketch of a network of pyramidal neurons, where a binary context signal (A or B) is projected on the apical distal compartment. Given the same sensory input, the target output changes accordingly to the context.
\textbf{B}. (Left) The network is able to reproduce the correct output trajectory even if the context is provided only in the first time steps (``turnoff'' experiment). Only one component of the 3D trajectory is represented for each context/target pair. (Right) An alternative model in which the context is projected on the basal compartment is no longer able to reproduce the correct output trajectory after the turnoff.
\textbf{C}. Summary of performances of the two model versions: context projected on apical (purple) vs basal (green) compartment during turnoff test in the middle of the trajectory. Mean squared error in the second part of the trajectory (no context) is compared with respect to error in the first part (context still active); mean and standard deviation are intended over 10 independent network/target realizations (round markers). $\mathrm{mse}$ between the output and the trajectory selected by the other context is also reported, as a reference (square markers).
\textbf{D}. 
(Left) Even in presence of a noisy context signal (top inset, $\sigma = 0.2$), the desired trajectory A (black) is reliably reproduced in output. (Right) When using a noisy basal context (top inset, $\sigma = 0.2$), at variance, the desired target A can no longer be reproduced by the network.
\textbf{E}. Average performances of the apical/basal context (purple/green, respectively) as a function of the noise standard deviation $\sigma$. Solid lines: $\mathrm{mse}$ between the output and the selected target trajectory. Dashed lines: $\mathrm{mse}$ between the output and the trajectory selected by the other context. Averages and error bars are intended over 10 independent network/target realizations.
}
\label{fig4}
\end{figure}

The distal apical compartment is designed not only to receive teaching signals, but also contextual information from other areas of the cortex, acting as a hint for the task to address. With this idea in mind, in this section we show that it is possible to exploit different context signals (projected through a Gaussian random matrix with variance $\sigma_\mathrm{cont}^2$) to flexibly select and recall one of the trajectories previously stored in the network.

In the simplest configuration, two different contexts, A and B, can be modeled through 2D time-constant binary signals projected on the distal apical compartment, $\bm{\chi}_{(1)} = (1,0)$ for A and $\bm{\chi}_{(2)} = (0,1)$ for B (Fig.~\ref{fig4}A).

During the training, each context is associated with a well defined target to learn (again a 3D trajectory, as defined in the previous section). In Fig.~\ref{fig4}B, left side, they are reported in red and black, respectively (only one of the three trajectories for each target is shown, for simplicity); same color-coding is used for associated context signals. To stabilize the learning, we exploited the trick of halving the learning rates $\eta$ and $\eta_{\mathrm{out}}$ every $100$ training iterations. The orthogonality of the contexts and related targets is further stressed by imposing a sparsification (of $75\%$ in the present case) in the random matrices we use to project the context and the target on the apical compartments of the network.

During the recall phase, the teacher signal is no longer present, while the context signal suggests to the network which of the learned trajectory to reproduce.
We show that when the context is projected to the network, the desired output is correctly recalled (Fig.~\ref{fig4}B, left side). Moreover, if the context signal is turned off in the middle of the trajectory, the network is still able to self-sustain its inner dynamics, thanks to recurrent connections (Fig.~\ref{fig4}B, left side), and correctly replicate also the remaining part of the selected trajectory. In Fig.~\ref{fig4}C we reported mean squared errors (averaged over 10 realizations), measured against both the correct target trajectory (purple square markers) and the wrong one (purple round markers), i.\,e. the one corresponding to the other context signal, both before turnoff and after it. In other words, the context works here as a ``suggestion'', so that once started the reproduction of the correct output trajectory, the context itself becomes useless.

To demonstrate the importance to project the context signal in the apical compartments, we compare these results with the case in which the context is projected in the basal ones (both during the training and the retrieval phases). In this case, the desired trajectory is correctly retrieved only when the context is on (Fig.~\ref{fig4}B, right side). However, in this case we observe that the basal context is interpreted as a necessary input, so that after the turnoff the network is no longer able to sustain bursts creation, in turn causing a dramatic drop in the retrieval performances (Fig.~\ref{fig4}B, right side). Corresponding average mean squared errors, again for both the correct target trajectory and the wrong one, both before turnoff and after it, are now reported in green in Fig.~\ref{fig4}C.

Furthermore, this apical context architecture is also robust against corruption in the context signal, which may be the case when at higher cortical levels there is only a mild preference in favor of which strategy to adopt (in comparison with the training phase, where each target is clearly and univocally associated with a sharp context signal). Here, a Gaussian white noise of variance $\sigma^2$ is added during test to context signals exploited in the training (Fig.~\ref{fig4}D, left panel, for $\sigma=0.2$). The produced trajectory is very similar to the trajectory selected by the context A (black dashed line) and different from the trajectory selected by the context B (red dashed line).
In Fig.~\ref{fig4}E, it is reported the average $\mathrm{mse}$ (average over 10 independent realizations of the experiment) between the output and the target trajectory (solid purple line) as a function of $\sigma$. As a reference, we also report the $\mathrm{mse}$ between the output and the trajectory selected by the other context signal (dashed purple line).
It is evident a resilience of the network with apical context, while the network with basal context suddenly loses the ability to reproduce the desired output already at low levels of noise (Fig.~\ref{fig4}D, right panel, for $\sigma=0.2$ and Fig.~\ref{fig4}E, green lines). At higher levels of noise, the basal context becomes in practice useless, while the apical one is still able to reproduce the target trajectory with a remarkably small error (Fig.~\ref{fig4}E).

\subsection{Hierarchical Imitation Learning}

\begin{figure}[h!]
\centering
\includegraphics[width=\textwidth]{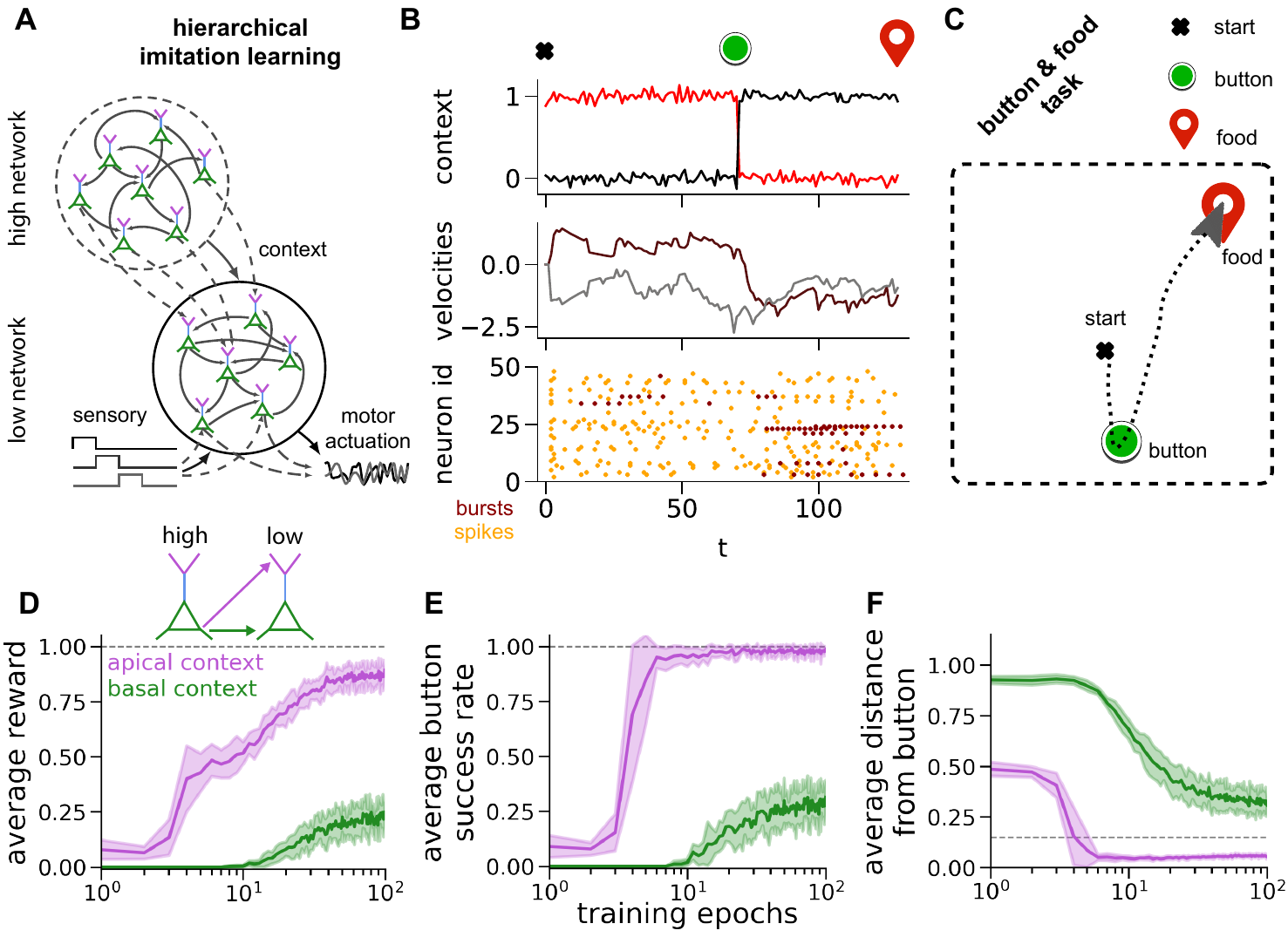}
\caption{
\textbf{Hierarchical Imitation Learning.}
\textbf{A}. A two-level network, where high-level neurons produce a signal that serves as a context for the neurons in the low-level network, allows implementing hierarchical policies. The two subnetworks receive two different but synchronized teaching signals in the training phase.
\textbf{B}. In the button \& food task, an agent placed at an initial position (black cross) in a 2D maze has to reach a button (green circle) in first place, so to unlock the food (red tag) and then reach for it. The high-level network chooses the order of the two subtasks ($\mathsf{reach\_button}$ and $\mathsf{reach\_food}$) and when to switch from one to the other. It projects the instruction as a contextual signal (top panel) to the apical compartments of the low-level network. The low-level network then produces the output (velocities of the agent, center panel) necessary to solve the subtask as a read-out of its internal bursting activity (bottom panel, brown dots; orange dots represent the spiking activity). 
\textbf{C}. A sample spatial trajectory produced by the agent after the training phase. Cross, green circle and red tag as in panel B.
\textbf{D}. Reward as a function of training epochs (average and standard error over 10 realizations, in lines and shadings respectively). Purple and green colors refer to the two different choices for the context projection on the low-network: apical or basal compartments, respectively (see also inset for a sketch of the two alternative models). Gray dashed line at $1.0$ indicates the maximum possible reward achievable.
\textbf{E}. Average success rate for pushing the button as a function of training epochs. $1.0$ is again the maximum possible value. Same color coding as in D.
\textbf{F}. Average distance from the button at the end of the episode. Gray dashed line represents the button size. Same color coding as in D.
}
    \label{fig5}
\end{figure}

The presence of an apical context that acts as a gating signal by flexibly selecting which dynamics to reproduce (and when), can be used as a building block for novel neural architectures that offer a biologically plausible implementation of \emph{hierarchical imitation learning} (HIL). In this work, we propose a two-level hierarchical network where the higher sub-module (high-network or ``manager'') computes the optimal strategy and exploits the context signal as a communication channel with the lower sub-module (low-network or ``worker''), which executes the selected task (Fig. \ref{fig5}A). We show that this architecture can efficiently solve the so-called \textit{button \& food} task.

In this scenario, an agent starts at the center of a square domain, which also features a button and an initially locked target (the ``food''). 
The button and the food are uniformly extracted on a unitary circle centered in the origin and in button position, respectively.

The global task effectively decomposes into two simpler sub-tasks or goals: $\mathsf{reach\_button}$ and $\mathsf{reach\_food}$. The high-network computes which (and when) of these two strategies to pursue, and communicates it to the low-network, which in turn implements the fine motor controls.
Both the high- and the low-network share the same input ($I = 80$ input units), the horizontal and vertical differences of both button and food positions with respect to the agent location, $\Delta^t =\{\Delta x_b^t, \Delta y_b^t,\Delta x_f^t, \Delta y_f^t\}$. Each of the $\Delta_i$ values is encoded by 20 input units with different Gaussian activation functions.

In this architecture, learning is implemented via a natural hierarchical extension of behavioral cloning: an expert system provides a collection of hierarchical demonstrations $\mathcal{D} = \left( \mathcal{D}_\mathrm{L}, \mathcal{D}_\mathrm{H} \right)$ for both sub-modules. A demonstration $\mathcal{D}_{\mathrm{L},\mathrm{H}}$ is a trajectory specification that can be described by the following tuple:
\begin{equation*}
    \mathcal{D}_{\mathrm{L},\mathrm{H}} = \left( \mathrm{state}_{\mathrm{L},\mathrm{H}}^t, \mathrm{action}_{\mathrm{L},\mathrm{H}}^t, \mathrm{goal}_{\mathrm{L}}^t \right), \quad t \in \left\{1, \dots, T \right\} 
\end{equation*}

\noindent where the $\mathrm{goal}^t$ component is missing for the high-network as it sits at the top of the hierarchy. The $\mathrm{state}_{\mathrm{L},\mathrm{H}}^t = \Delta^t$ component, shared between the two sub-modules, is a description of the agent position in the environment.
The high-network output $\mathrm{action}_\mathrm{H}^t$ encodes the selected strategy, and it is projected to the low-network as a contextual signal in the distal apical compartment, where it is interpreted as the low-network $\mathrm{goal}_\mathrm{L}^t$. We define this target behavior to be the $2D$ signal:


\begin{equation*}
\bm{y}^{\star, t}_\mathrm{H} = \bm{\chi}_{(1)} \Theta \left(t < t_{\lock} \right) + \bm{\chi}_{(2)} \Theta \left(t > t_{\lock} \right)\,,
\end{equation*}
where $\bm{\chi}_{(1)}=(1,0)$ and $\bm{\chi}_{(2)}=(0,1)$, and $t_{\lock}$ is the time when the button is reached (Fig.~\ref{fig5}B, top). Intuitively, it selects the $\mathsf{reach\_button}$ sub-policy for the first part of the task and then switches to $\mathsf{reach\_target}$, once the latter has been unlocked.

Given the input $\mathrm{state}^t_\mathrm{L}$ and the context $\mathrm{goal}^t_\mathrm{L}$, the low-network 
is tasked to reproduce as output $\mathrm{action}^t_\mathrm{L}$ the velocity vector $\bm{y}^{\star, t}_\mathrm{L} = \bm{v}^t = (v_x^t, v_y^t)$, where velocity components are computed so to reach the selected target in a straight line (Fig.~\ref{fig5}B, center). Both high- and low-network outputs are computed as linear read-outs of their internal bursting activities, as described in Section~\ref{subsec:target_based} (Fig.~\ref{fig5}B, bottom, for the low-network).

We implement the cloning procedure in a supervised fashion, following the same procedure as the one described in Section \ref{subsec:target_based}: the two sub-modules are trained to reproduce their target outputs given their set of inputs (context included).

Finally, the two-layer network is tested in closed-loop in the environment described above (Fig.~\ref{fig5}C). The performances are measured via the following quantity:
\begin{equation*}
    \rho = \frac{\Xi_{\lock} \, r_0}{\min_{t > t_{\lock}} d \left(\bm{x}^t_\mathrm{agent}, \bm{x}_\mathrm{food} \right)} \, ,
\end{equation*}

\noindent where $r_0$ is the button and food size, $\Xi_{\lock}$ is the button-state indicator variable (0 when the button is locked and 1 otherwise), and finally $d \left(\bm{x}^t_\mathrm{agent}, \bm{x}_\mathrm{food} \right)$ is the Euclidean distance between the agent and food positions at time $t$. The condition for a successful button-press (a switch from locked to unlocked state) and target-reach is taken to be $d \left(\bm{x}^t_\mathrm{agent}, \bm{x}_\mathrm{btn/food} \right) \le r_0$. Note how this choice effectively prevents the apparent divergence in the expression for $\rho$ as the episode is stopped when the target is reached, thus inducing a theoretical maximum achievable score of $\rho_\mathrm{max} = 1$. Otherwise, $0<\rho<1$ if the button has been unlocked, but the food has not been reached within the assigned time window, or $\rho=0$ if the button has not been reached at all.

After the presentation of many randomly positioned button-food pairs, we observe that the hierarchical two-level network learns to correctly and efficiently solve the button \& food task, with an average final score $\rho = 0.88 \pm 0.04$ and over $70\%$ of success rate (i.e., both button-press and target-reach conditions were met).
In Fig.~\ref{fig5}D, purple line, we report the average reward (over 10 independent realizations) as a function of training epochs.
Similarly, in Figs.~\ref{fig5}E-F, purple curves, it is reported the average success rate in pushing the button, and the average minimum distance from the button, respectively.

We run an additional experiment, where the high-network output is projected to the basal compartment of the low-network (rather than to the apical one, see Fig.~\ref{fig5}D, top inset). The results (averaged over 10 further independent realizations) are reported in Figs.~\ref{fig5}D-F, green lines. This choice leads to poor performances of the hierarchical policy ($\rho = 0.24 \pm 0.07$), demonstrating, also in this case, the necessity of a contextual signal of a different nature with respect to somatic input signals.

\section{Methods}

\subsection{The model}

We defined a neuron model, inspired to the pyramidal L5 neuron, composed of three different compartments: a basal one ($b$) and two apical ones, named proximal ($p$) and distal ($d$), respectively (see Fig.~\ref{fig1}B for reference). All the model parameters are reported in Table~\ref{table1}.

If we focus on a particular neuron $i$, with $i = 1, \dots, N$, its membrane potential vector $\bm{\mathrm{v}}^t_i = \left(v^t_i, \, u^t_i, \, u_i^{\star, t} \right)$ (the membrane potentials of basal, proximal apical, and distal apical compartments, respectively) follows a leaky-integrate-and-fire dynamics, which we can generically be written as:

\begin{equation}
    \bm{\mathrm{v}}_i^{t+1} = \left[\left(1 - \frac{dt}{\tau_m} \right) \bm{\mathrm{v}}_i^t + \frac{dt}{\tau_m} \bm{\mathrm{I}}_i^{t+1} \right] \left(1 - \bm{\mathrm{s}}_i^t \right) + \bm{\mathrm{v}^{\circlearrowright}}  \bm{\mathrm{s}}_i^t \, ,
\label{EQ:Vector_Membrane_Potential_Dynamics}
\end{equation}

\noindent where the vector quantities $\bm{\mathrm{I}}^t_i = (I_{(b), i}^t,  \, I_{(p), i}^t, \, I_{(d), i}^t )$, $\bm{\mathrm{s}}_i^t = (z_i^t, \, a_i^t, \, a_i^{\star, t})$ and $\bm{\mathrm{v}^{\circlearrowright}} = (\mathrm{v}^\circlearrowright_{(b)}, \, \mathrm{v}^\circlearrowright_{(p)}, \, \mathrm{v}^\circlearrowright_{(d)})$ represent the input current, the neuron spike and the reset potential, respectively, for each compartment (see following sections for their explicit definitions). The membrane potential vector $\bm{\mathrm{v}}_i^t$ defines the stochastic emission of the spike at the subsequent time step $\bm{\mathrm{s}}_i^{t+1}$ via a sigmoid probability:

\begin{equation}
    p \left(\bm{\mathrm{s}}_i^{t+1} | \bm{\mathrm{v}}_i^t \right) = \frac{\exp \left[\bm{\mathrm{s}}_i^{t+1} \left(\frac{\bm{\mathrm{v}}_i^t - v_\mathrm{thr}}{\delta v} \right) \right]}{1 + \exp \left(\frac{\bm{\mathrm{v}}_i^t - v_\mathrm{thr}}{\delta v} \right)} \, ,
\label{EQ:Vector_Probability_Sigmoid_Definition}
\end{equation}

\noindent
with $v_{\mathrm{thr}}$ being the firing threshold for the membrane potential and $\delta v$ a model parameter controlling the probabilistic nature of the firing process (both assuming the same value for the three compartments, in our model). In the $\delta v \to 0$ limit, the spike-generation rule \eqref{EQ:Vector_Probability_Sigmoid_Definition} becomes deterministic and reduces to:

\begin{equation*}
p(\bm{\mathrm{s}}^{t+1}|\bm{\mathrm{v}}^t)=\Theta[\bm{\mathrm{s}}^{t+1}(\bm{\mathrm{v}}^t-v_{\mathrm{thr}})] \, .
\end{equation*}

\noindent
We remark that in all the numeric implementation of model dynamics we assume the deterministic dynamics ($\delta v\rightarrow0$).

\subsubsection{Temporal filtering and windows}
\label{subsubsec:temporal_filtering}

We introduce the exponential filtering function $\mathsf{filter}\left(\xi^t, \tau \right)$, defined recursively as:
\begin{equation}
    \mathsf{filter} \left( \xi^{t+1}, \tau \right) \equiv \exp \left(-\frac{dt}{\tau} \right) \mathsf{filter} \left( \xi^t, \tau \right) + \left( 1 - \exp \left(-\frac{dt}{\tau} \right) \right) \xi^{t+1} \, .
\label{EQ:Filtering_Function}
\end{equation}
Basal spike signals are time-filtered through suitable time constants, depending on the direction they propagate. Using the previous definition, we introduce the following time-filtered quantities:
\begin{align}
\hat{z}_{i}^{t+1} &\equiv \mathsf{filter} \left(z_{i}^{t+1}, \tau_s \right) \\
\hat{z}_{\mathrm{ro},i}^{t+1} &\equiv \mathsf{filter} \left(z_{i}^{t+1}, \tau_\mathrm{ro} \right) \\
\hat{z}_{\mathrm{soma},i}^{t+1} &\equiv \mathsf{filter} \left(z_{i}^{t+1}, \tau_\mathrm{targ} \right)
\end{align}
Such filtering is also applied to the adaptation term $\omega_i^t$ contributing to the input current of the basal compartment, which is time-smoothed as:

\begin{equation}
    \omega_i^{t+1} \equiv \mathsf{filter} \left(z_{i}^{t+1}, \tau_\omega \right).
\end{equation}

\begin{table}[t!]
    \caption{\textbf{Parameter of numerical simulations}. Many parameters have the same value for all the simulations reported in the main text figures (columns on the right). When not the case, the different values used are clearly indicated (columns on the left). For Fig.~\ref{fig5}, two values for the low-network (L) and the high-network (H), respectively, have been reported, when different from each other. For $\eta$ and $\eta_{\mathrm{out}}$ for Fig.~\ref{fig4}, we report the initial parameter values, as during learning they are discounted (as discussed in Section~\ref{subsec:context_selection}).}
    \label{table1}

\vskip 0.15in
\begin{center}
\begin{small}
\begin{sc}
\setlength{\tabcolsep}{5pt}
\begin{tabularx}{\textwidth}{
    p{0.05\textwidth}%
    >{\centering\arraybackslash}p{0.08\textwidth}%
    >{\centering\arraybackslash}p{0.08\textwidth}%
    >{\centering\arraybackslash}p{0.125\textwidth}%
    >{\centering\arraybackslash}p{0.085\textwidth}%
    p{0.08\textwidth}%
    >{\centering\arraybackslash}p{0.130\textwidth}%
    p{0.06\textwidth}%
    >{\centering\arraybackslash}p{0.130\textwidth}%
}
\toprule
\multicolumn{4}{c}{Figure-specific parameters} & & \multicolumn{4}{c}{\multirow{2}{*}{Universal parameters}} \\
 & Fig. 3 & Fig. 4 & Fig. 5 [L\,--\,H] & & \multicolumn{4}{c}{} \\
\midrule
$N$                      &  500   &  1000  &  500\,--\,500 & & $\tau_{\mathrm{m}}$ & 20 $\mathrm{(ms)}$ & $u_0$ & -6 $\mathrm{(mV)}$ \\
$N_e$                    &  400   &  800   &  400\,--\,400 & & $\tau_{\mathrm{s}}$ & 2 $\mathrm{(ms)}$ & $u_0^{\star}$ & -6 $\mathrm{(mV)}$ \\
$N_i$                    &  100   &  200   &  100\,--\,100 & & $\tau_{\mathrm{out}}$ & 10 $\mathrm{(ms)}$ & $v_\mathrm{thr}$ & 0 $\mathrm{(mV)}$ \\
$\sigma_{\mathrm{targ}}$ &  20    &  30    &  0\,--\,100   & & $\tau_{\mathrm{targ}}$ & 20 $\mathrm{(ms)}$ & $\vartheta_{\mathrm{soma}}$ & $2.5 \times 10^{-2}$ \\
$\sigma_{\mathrm{in}}$   &  12    &  12    &  20           & & $\tau_{\omega}$ & 200 $\mathrm{(ms)}$ & $\vartheta_{\mathrm{burst}}$ & $1.25 \times 10^{-2}$ \\
$\eta$                   &  10    &  10    &  0\,--\,0.25  & & $b$ & 100 \\
$\eta_{\mathrm{out}}$    &  0.01  &  0.01  &  0.03         & & $v_{\mathrm{reset}, b}$ & -20 $\mathrm{(mV)}$ \\
$I$                      &  5     &  50    &  n.d.         & & $v_{\mathrm{reset}, d, p}$ & -160 $\mathrm{(mV)}$ \\
$\sigma_{\mathrm{cont}}$ &  0     &  20    &  50\,--\,0    & & $v_0$ & -1 $\mathrm{(mV)}$ \\
\bottomrule
\end{tabularx}
\end{sc}
\end{small}
\end{center}
\vskip -0.1in
\end{table}

The occurrence of a somatic spike opens a temporal somatic window:
\begin{equation}
\shat{z}_{i}^{t} = \Theta[\hat{z}_{\mathrm{soma},i}^{t} - \vartheta_{\mathrm{soma}}] \, .
\end{equation}

The onset of a burst in the proximal or distal compartments is induced by the coincidence between the somatic window $\shat{z}_{i}^{t}$ and an apical proximal $a_i^t$ or an apical distal $a_i^{\star, t}$. The bursts can hence be expressed, respectively, as:
\begin{align}
B_{i}^{t+1} &= \shat{z}_i^{t}  a_{i}^{t+1}\\
B_{i}^{\star, t+1} &= \shat{z}_i^{t} a_{i}^{\star, t+1}
\label{eq:B_i_star}
\end{align}
Aiming for a time-window variable that is active during burst activity, we can iterate the same construction developed for spikes and consider the filtered burst-onset $\hat{\bm{\mathrm{B}}}_i^t$:
\begin{align}
\hat{B}_{i}^{t+1} &\equiv \mathsf{filter} \left( B_i^{t+1}, \tau_\mathrm{targ} \right)\\
\hat{B}_{i}^{\star, t+1} &\equiv \mathsf{filter} \left( B_i^{\star, t+1}, \tau_\mathrm{targ} \right)
\end{align}
One can again use these time-filtered quantities to introduce proximal and distal burst windows as:
\begin{align}
\shat{B}_{i}^{t+1} &= \Theta[\hat{B}_{i}^{t+1} - \vartheta_{\mathrm{burst}}]\\
\shat{B}_{i}^{\star, t+1} &= \Theta[\hat{B}_{i}^{\star, t+1} - \vartheta_{\mathrm{burst}}]
\end{align}
When at least one among proximal and distal bursts is above threshold, we finally have a neural burst activity window:
\begin{equation}
    \shat{B}_{\lor,i}^{t+1} = \shat{B}_{i}^{t+1} \lor \shat{B}_{i}^{\star, t+1} \, ,
\end{equation}
which is the quantity that will appear in the dynamics of the compartments.


\subsubsection{Basal compartment}

The membrane potential of the basal compartment evolves following the equations:
\begin{align*}
v_{i}^{t+1} &= \left[\left(1-\frac{dt}{\tau_m}\right)v_{i}^t + \frac{dt}{\tau_m}I_{(b),i}^{t+1}\right](1 - z_{i}^t) + \mathrm{v}^{\circlearrowright}_{(b)} z_{i}^t \\
I_{(b),i}^t &= \underbrace{\sum_{j=1}^N w_{ij}^{b\to b} \hat{z}_{j}^t}_{\substack{\mathsf{recurrent\   basal-basal}\\ \mathsf{connections}}} + \underbrace{\sum_{k=1}^{n_{\mathrm{inp}}} w_{ik}^{\mathrm{inp}} I_k^{\mathrm{inp},t}}_{\substack{\mathsf{sensorial}\\ \mathsf{input}}} + \underbrace{\beta \shat{B}_{\lor, i}^t}_{\mathsf{\substack{\mathsf{extra\ current}\\ \mathsf{from\ coincidence}}}} - \underbrace{b \,\hat{\omega}_i^t}_{\mathsf{\substack{\mathsf{adaptation}\\ \mathsf{term}}}} + v_0
\end{align*}
where recurrent connections targeting the basal compartments ($w_{ij}^{b\to b}$) are in fact set to zero in our model. The contribution of sensorial input is given by the input current $I_k^{\mathrm{inp}, t}$, randomly projected to the neurons through the weights $w_{ik}^\mathrm{inp}$, while $v_0$ is a compartment-specific constant input.

Finally, to induce a high frequency burst  during the burst window $\shat{B}_{\lor, i}^t$, we introduce the extra current term$ \beta \shat{B}_{\lor, i}^t$  (we set $\beta=20$). Also, the basal reset potential is suitably increased during the burst window:
\begin{equation*}
    \bm{\mathrm{v}}^{\circlearrowright}_{(b)} = \frac{v_{\mathrm{reset}, b}}{1 + \alpha \shat{B}_{\lor, i}^{t}} \, ,
\end{equation*}
where $v_{\mathrm{reset}, b}<0$ is a compartment-specific scalar and $\alpha$ is a constant model parameter (we set $\alpha=2$).


\subsubsection{Apical proximal compartment}
The apical proximal compartment of each neuron is connected to basal compartments of all the neurons through recurrent connections $w_{ij}^{b \to p}$. These recurrent connections are the object of the training procedure and are adjusted to reproduce the desired target. The equations for this compartment dynamics are:

\begin{align*}
u_{i}^{t+1} &= \left[\left(1-\frac{dt}{\tau_m}\right) u_{i}^t + \frac{dt}{\tau_m}I_{(p),i}^{t+1}\right](1 - a_{i}^t) + \mathrm{v}^\circlearrowright_{(p)} a_{i}^t\\
I_{(p),i}^t &= \underbrace{\sum_{j=1}^N w_{ij}^{b \to p} \hat{z}_{j}^t (t)}_{\substack{\mathsf{recurrent\   basal-proximal}\\ \mathsf{connections}}} +\ u_0
\end{align*}
The reset potential for the proximal apical compartment $\mathrm{v}^\circlearrowright_{(p)} = v_{\mathrm{reset}, p}$ is a compartment-specific scalar, independent of burst activity, while $u_0$ is the compartment-specific constant input.

\subsubsection{Apical distal compartment}

The signal to be learned (target) is considered as an input for the apical distal compartment: during the learning stage it is injected via a projection matrix $w_{ik}^\mathrm{targ}$, while it is completely absent during spontaneous activity. The coefficient $f_\mathrm{apic} \in \left\{0, 1\right\}$ is used to gate this stage transition.

Also, the input from the context (again randomly projected on the $N$ neurons via the $w_{ik}^\mathrm{cont}$ matrix) is given as input for the apical distal compartment. The equations for the apical distal compartment then read:

\begin{equation*}
u_{i}^{\star, t+1} = \left[\left(1-\frac{dt}{\tau_m}\right) u_{i}^{\star, t} + \frac{dt}{\tau_m} I_{(d),i}^{t+1}\right](1 - a_{i}^{\star, t}) + \mathrm{v}^{\circlearrowright}_{(d)} a_{i}^{\star, t}
\end{equation*}

\begin{align*}
I_{(d),i}^t &= \underbrace{f_\mathrm{apic}\sum_{k=1}^{n_{\mathrm{output}}} w^{\mathrm{targ}}_{ik}y_k^{\star, t}}_{\mathsf{target/teach\ input}} + \underbrace{\sum_{k=1}^{n_{\mathrm{cont}}}w_{ik}^{\mathrm{cont}}C_k^t}_{\mathsf{context\ input}} +\  u^\star_0 
\end{align*}

\noindent
where $y_k^{\star, t}$ is the target signal and $C_k^t$ is the context signal, while $u_0^\star$ is the compartment-specific constant input.

\subsection{Derivation of the learning rule}

We formulate the learning process as the maximization of the probability of observing the desired spatio-temporal pattern of bursts. By expressing such probability in terms of the recurrent network connections, we obtain an explicit expression for the learning rule. In doing so, we directly extend previous approaches used for learning the target pattern of spikes \cite{pfister2006optimal,rezende2014stochastic,gardner2016supervised,muratore2021target}.

We start by writing the probability to produce a burst in the neuron $i$ at time $t$, given the somatic window $\shat{z}_i^t$. We propose the following compact formulation:

\begin{equation}
    p(B_{i}^{\star, t+1}| \shat{z}_{i}^t) = \frac{\exp{\left[B_{i}^{\star, t+1} \Phi_{i}^t (\shat{z}_{i}^t) \right]}}{1+\exp{\left[ \Phi_i^t(\shat{z}_{i}^t) \right] } } \, ,
\end{equation}

\noindent
where we have introduced $\Phi_{i}^t (\shat{z}_{i}^t) = a_i^{t+1} \shat{z}_{i}^t/\delta v - (1-\shat{z}_{i}^t)\gamma$. By definition, a burst can only happen by means of a basal-apical spike coincidence, represented by the $a_i^{t+1} \shat{z}_i^t$ term; when the basal window is open ($\shat{z}_i^t = 1$), the burst probability reduces to the usual sigmoidal function. When the window is closed and $\shat{z}_i^t = 0$, we have $\Phi_i^t \left( \shat{z}_i^t \right) = -\gamma$; we can thus tune the $\gamma$ parameter to model the burst probability. In practice, we work in the $\gamma \to \infty$ limit where $\lim_{\gamma \to \infty} p (B_i^{\star, t + 1} | \shat{z}_i^t = 0) = 0$, which agrees to the intuitive understanding that a closed basal window prevents any burst activity.

We introduce the log-likelihood $\mathcal{L}$ of observing a given target burst activity $\bm{\mathrm{B}^\star}$ given the basal-to-proximal connections $w_{ij}^{b \to p}$ as:

\begin{equation}
\mathcal{L} \left( \bm{\mathrm{B}^{\star}} | w^{b \to p} \right) = \sum_{i,t} \left[ B_{i}^{\star, t+1} \Phi_{i}^t (\shat{z}_{i}^t) \right. - \log{ \left( 1 + \exp{\left[ \Phi_i^t(\shat{z}_{i}^t) \right] } \right)} \Big] \, .
\end{equation}

\noindent
We can then maximize this likelihood by adjusting the synaptic connections so to achieve the target burst activity $\bm{\mathrm{B}^\star}$. By differentiating with respect to the recurrent apical weights, we get:
\begin{equation}
\frac{\partial \mathcal{L}( \bm{\mathrm{B}}^{\star} | w^{b \to p})}{\partial w_{ij}^{b \to p}} =  \left[B_{i}^{\star, t+1} -p( B_{i}^{t+1}=1) \right] \shat{z}_{i}^t e_j^t \, ,
\end{equation}
where we have introduced the following two quantities:
\begin{equation*}
p( B_{i}^{t+1}=1) =\frac{\exp{\left[ \Phi_{i}^t (\shat{z}_{i}^t) \right]}} {1+\exp{\left[ \Phi_i^t(\shat{z}_{i}^t) \right]}} \quad \mathrm{and} \quad e_j^t = \frac{\partial u_i^t}{ \partial w_{ij}^{b \to p}} \, ,
\end{equation*}

\noindent
respectively a sigmoid probability for $\Phi_{i}^t$ and the \textit{spike response function} \cite{urbanczik2014learning}.

Given the basal window state $\shat{z}_i^t$, the target burst sequence is uniquely defined by the input projected to the apical distal compartment and can be written as $B_i^{\star, t+1} = \shat{z}_i^t a_{i}^{\star, t+1}$.
If we take the deterministic limit of the model ($\delta v \rightarrow 0$, where $p (B_i^{t+1} = 1) = a_i^{t+1} \shat{z}_i^t)$ and then note that $\shat{z}_i^t \, \shat{z}_i^t = \shat{z}_i^t$, we can rewrite the previous expression in a cleaner form:
\begin{equation}
\frac{\partial \mathcal{L} ( \bm{\mathrm{B}^\star}  | w^{b \to p})}{\partial w_{ij}^{b \to p}} =  \left[a_{i}^{\star, t+1} - a_{i}^{t+1} \right] \shat{z}_{i}^t e_j^t \, .
\end{equation}
This means that the spikes in the proximal apical compartment $a_{i}^{t+1}$ should mimic the ones in the distal one $a_{i}^{\star, t+1}$ when the somatic window $\shat{z}_i^t$ is open, in order to maximize the probability that bursts generated by the network reproduce the target pattern of bursts.

For simplicity, we discussed this version of the learning rule. However, in this work we actually exploited the non-deterministic version of the plasticity rule (finite $\delta v = 0.1$),
that can be rewritten as:
\begin{equation}
\frac{\partial \mathcal{L} ( \bm{\mathrm{B}^\star}  | w^{b \to p})}{\partial w_{ij}^{b \to p}} =  \left[a_{i}^{\star, t+1} - p \left( a_i^{t+1}=1 | u_i^t \right) \right] \shat{z}_{i}^t e_j^t \, ,
\end{equation}

\noindent
where:

\begin{equation}
p \left(a_i^{t+1}=1 | u_i^t \right) = \frac{\exp   \left(\frac{u_i^t - v_\mathrm{thr}}{\delta v} \right)  }{1 + \exp \left(\frac{u_i^t - v_\mathrm{thr}}{\delta v} \right)} \, .
\end{equation}


We stress here how in the derivation we considered the basal-window state $\shat{z}_i^t$ as given. Consequently, the target burst sequence $\bm{\mathrm{B}^\star}$ is uniquely defined by the input projected to the apical distal compartment and the likelihood is well-defined. Though we are aware of the feedback influence of the burst activity on the basal-window configuration (bursts induce basal spikes, see the equation for basal current $I_{(b), i}^t$ in the \textsc{basal compartment} section), we chose to neglect such contribution as it would have severely increased the difficulty of the derivation. The convergence to the chosen target thus cannot be guaranteed. 

However, despite this limitation, it is important to note that the pattern of apical spikes $\{a^{\star}\}$ does not change during learning: it is entirely determined by the original teaching signal $y ^{\star}$ and the variance $\sigma_{\rm targ}$ of its random projection to the network. As target bursts only occur after coincidence of an apical spike $a^{\star}$ and a basal spike $z$, the pattern $\{B^{\star}\}$ is necessarily a subset of the fixed distal apical spikes $\{a^{\star}\}$, and thus cannot diverge. In principle, it is still possible that the target pattern oscillates between slightly different subsets of $\{a^{\star}\}$. In practice, we do not observe such behavior and provide a numerical demonstration that the target pattern of bursts converges to a well-defined pattern (see Appendix for details).

\section{Discussion}

It is more and more evident that dendrites are capable of producing spikes \cite{gasparini2004initiation} and performing complex and non-linear computations \cite{poirazi2020illuminating}.
A famous example is the capability to initiate a broad calcium action potentials (``Ca$^{2+}$ spikes'') near the apical tuft of pyramidal layer-5 neurons, that produces a long (up to 50 ms in vitro) plateau-type depolarization \cite{larkum2013cellular}. The coincidence between this phenomenon and a somatic spike induces high-frequency somatic bursts during such depolarization.
In the present work, we model such mechanism through the variable $\shat{B}_i^t$ (see Eq.~\eqref{eq:B_i_star}), that is $1$ for a 30 ms time window, after the coincidence between the apical ($a_i^t$) and the somatic spike ($z_i^t$).

We show that this mechanism enables pyramidal neurons to naturally support target-based learning, that is easily applicable to, e.g., store and recall tasks. Moreover, it makes possible to use contextual signals to flexibly select the desired output from a repertoire of learned dynamics, acting as a hint or suggestion.

Also, we argue that this framework provides a natural solution to a general problem in learning: the plasticity-stability dilemma. A neural network requires to quickly capture statistical irregularities and learn new information, still retaining network stability to prevent forgetting previous memories.
A first instantiation of this problem is Hebbian plasticity, that provides a positive feedback loop and leads to unstable runaway activity \cite{abbott2000synaptic}. 
To resolve this, it has been suggested that homeostatic processes keep the network activity stable \cite{turrigiano2004homeostatic}.
In \cite{wilmes2022} the authors show that the gating of plasticity in dendrites can improve network stability without compromising plasticity.
This problem was also faced in a network of bursting neurons, in \cite{vercruysse2021self}, where the authors define a homeostatic rule to regulate the bursting and the firing activity to a target value. 
We argue that our target-based approach intrinsically solves the  plasticity-stability dilemma.
Indeed, the learning instability is mitigated by an intrinsic separation of timescales between the proposed target (that changes slowly in time, incrementing stability and preventing activity runaway), and the quick changes in synaptic weights (allowing a quick learning of proposed targets).

The neuronal architecture we propose allows to build networks with hierarchical architectures, which in turn are suited to orchestrate \textit{hierarchical imitation learning} (HIL) \cite{le2018hierarchical}. It enables the decomposition of challenging long-horizon decision-making tasks into simpler sub-tasks, improving both learning speed and transfer learning, as skills learned by sub-modules can be re-used for different tasks. In our work, a high-level network (the manager) selects the correct policy for the task, suggesting it as a contextual signal to the low-level network (the worker), in charge of actually executing it. We also show how considering contextual information as an input for the apical compartment (instead of the basal one) is crucial for the correct decomposition and accomplishment of the task, in agreement with the biological interpretation of apical dendritic inputs as contextual signals from other cortical areas.

Though our hierarchical imitation learning approach requires devising handcrafted solutions for the different layers, our message is that the architecture we propose can efficiently support the implementation of hierarchical policies.
In future works, we plan to replace behavioral cloning with more general learning schemes, such as feudal networks \cite{vezhnevets2017feudal}, where the ``high network'' (manager) moves in a latent space and the ``low network'' (worker) translates it into meaningful behavioral primitives.

To our knowledge, no other existing works propose a biologically plausible architecture to implement HIL 
Furthermore, our model prepares the ground for further biological explorations. Tuning model parameters (e.g., the adaptation strength $b$) allows simulating the transition between different brain states (e.g., sleep and awake) \cite{wei2018differential,goldman2020brain,tort2021attractor}. Possible future investigation topics include the replay of patterns of bursts during sleep \cite{kaefer2020replay}, and the effect of sleep on tasks performances \cite{wei2018differential,capone2019sleep,golosio2021thalamo,capone2022towards}.

\section*{Source code availability}
The source code is available for download under CC-BY license in the \href{https://github.com/cristianocapone/LTTB}{https://github.com/cristianocapone/LTTB} public repository.

\section*{Acknowledgements}
This work has been supported by the European Union Horizon 2020 Research and Innovation program under the FET Flagship Human Brain Project (grant agreement SGA3 n. 945539 and grant agreement SGA2 n. 785907) and by the INFN APE Parallel/Distributed Computing laboratory.

\printbibliography

\newpage
\appendix

\renewcommand{\thefigure}{S\arabic{figure}}
\setcounter{figure}{0}
\setcounter{section}{0}

\onecolumn

\begin{center}
\rule{\textwidth}{0.05cm}

\Large \textbf{Appendix\\ Beyond spiking networks: the computational advantages of dendritic amplification and input segregation}
\rule{\textwidth}{0.05cm}
\end{center}

\section*{Numerical evidence of convergence}

As mentioned in the main text, we can not provide a mathematical proof of the convergence toward the chosen target of burst activity by means of the learning rule proposed here. However, strong evidences in this direction can be found numerically.

We run several independent realizations of the same task of Fig.~\ref{fig3}, i.e., the store and recall of a 3D trajectory. We look at the distance between the target and the spontaneous spatio-temporal pattern of bursts during the training, and also at the self-distance in the pattern of target bursts across consecutive training iterations.

The parameters used for these simulations (when different from those used for Fig.~\ref{fig3} and reported in Table~\ref{table1}) are: $\eta=2.5$, $\eta_{\mathrm{out}}=2.5 \times 10^{-3}$, $\sigma_{\mathrm{targ}}$ variable from $10$ (black) to $1000$ 
(yellow). Data averaged over $10$ independent network/target realizations for each value of $\sigma_{\mathrm{targ}}$.

The distance between two patterns of bursts $A =\{A_i^t\}$ and $B = \{B_i^t\}$ is defined as the root mean squared error of their element-wise difference:
\begin{equation*}
    \mathcal{D}(A,B) \equiv \sqrt{\frac{1}{N\,T}\sum_{i,t} \left(A^t_i-B_i^t\right)^2} \, .
\end{equation*}

\begin{figure}[b!]
\centering
\includegraphics[width=\textwidth]{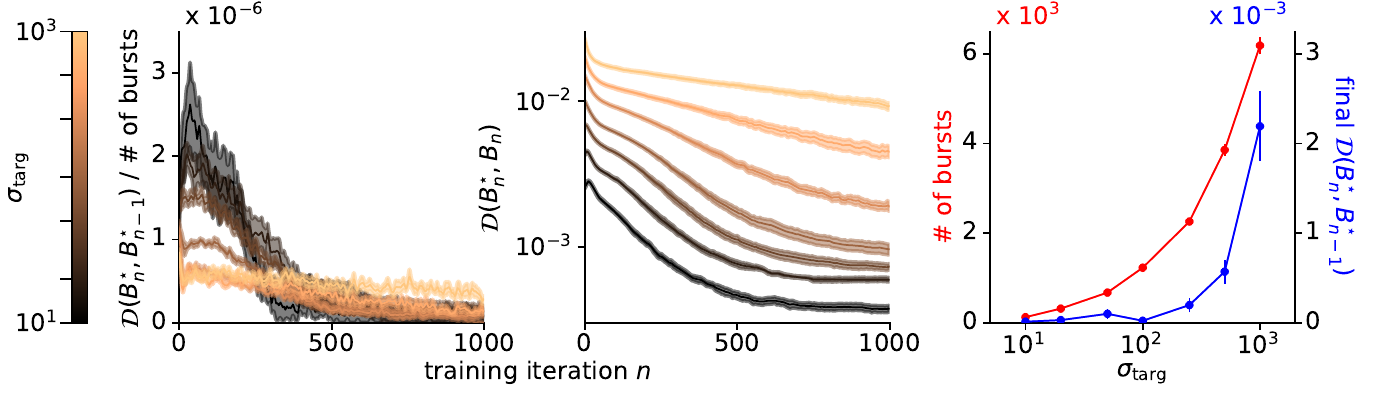}
\caption{
\textbf{Convergence of the target pattern of bursts vs $\sigma_{\mathrm{targ}}$.}
(Left) Self-distance of the target pattern of bursts $\mathcal{D}(B^{\star}_n,B^{\star}_{n-1})$, normalized by the total number of bursts, as a function of the number~$n$ of learning iterations, for different  $\sigma_{\mathrm{targ}}$ values (lower to higher values, from dark to light).
(Middle) Distance between the target and spontaneous pattern of bursts $\mathcal{D}(B^{\star}_n,B_{n})$, again as a function of the number of learning iterations $n$ for different values of $\sigma_{\mathrm{targ}}$. Same color coding as in the first panel. 
(Right) Average self-distance of the target pattern of bursts at the end of the training (blue) and average number of target bursts (red) as a function of $\sigma_{\mathrm{targ}}$.
}
\label{figs1}
\end{figure}

For small values of $\sigma_{\mathrm{targ}}$, comparable to the ones used for main text figures, target bursts rapidly settle after some hundreds of training iterations  (Fig.~\ref{figs1}, left); within the same training scale, also spontaneous burst activity matches the target one, with a negligible error (Fig.~\ref{figs1}, middle).

We prove that in a broad range of $\sigma_{\mathrm{targ}}$ values (roughly up to $\sigma_{\mathrm{targ}}=100$), the target pattern of bursts converges to a well-defined one (Fig.~\ref{figs1}, right, blue dots), while for higher values of $\sigma_{\mathrm{targ}}$ the convergence slows down. This is related to the increase in the number of bursts for high values of $\sigma_{\mathrm{targ}}$ (Fig.~\ref{figs1}, right, red dots).

\begin{figure*}[t!]
\centering
\includegraphics[width=\textwidth]{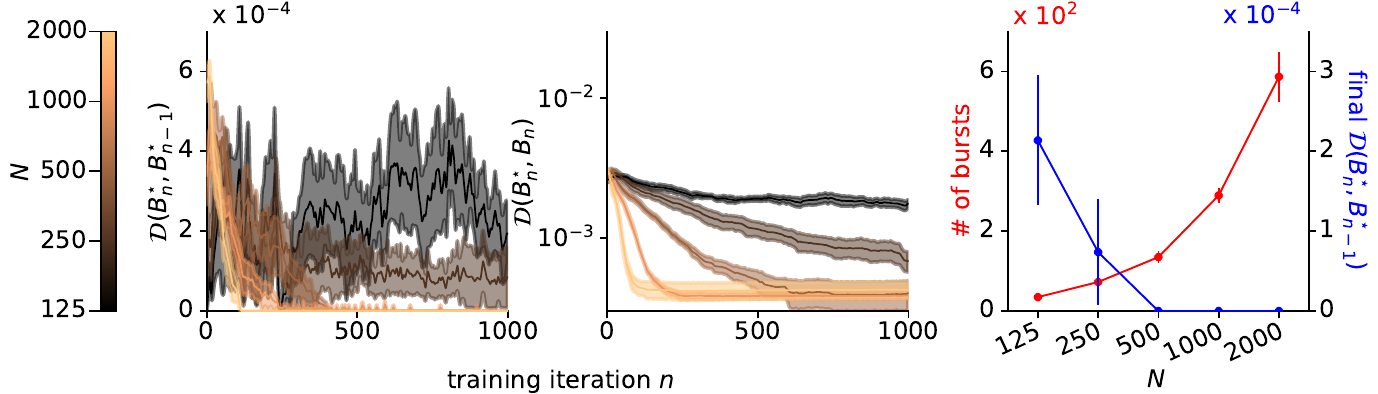}
\caption{
\textbf{Convergence of the target pattern of bursts vs $\mathbf{N}$.}
(Left) Self-distance of the target pattern of bursts $\mathcal{D}(B^{\star}_n,B^{\star}_{n-1})$ during the training, for different network sizes $N$ (lower to higher values, from dark to light).
(Middle) Distance between the target and spontaneous pattern of bursts $\mathcal{D}(B^{\star}_n,B_{n})$ after $n$ learning iterations, for the same network sizes as in the first panel (same color coding).
(Right) Average self-distance of the target pattern of bursts at the end of the training (blue) and average number of target bursts (red) as a function of $N$.
}
\label{figs2}
\end{figure*}

To eventually check also the dependence on the network size, we made further simulations for different sizes ($N=125, \, 250, \, 500, \, 1000, \, 2000$) at a given value $\sigma_{\rm targ}=10$ (10 independent network/target realizations for each value of $N$). The convergence at the end of the training for the same quantities in Fig.~\ref{figs1} is always observed (Fig.~\ref{figs2}, left and middle panels). Actually, such convergence is even faster the larger the networks, while the total number of bursts increases with $N$, as expected (Fig.~\ref{figs2}, right panel).

\end{document}